\documentclass[11pt,a4paper]{article}%
\usepackage{amsmath,amssymb,amsfonts,wasysym}
\usepackage{graphicx}
\usepackage{color}
\usepackage{float}
\usepackage[bf,footnotesize]{caption2}
\usepackage[]{hyperref}%
\numberwithin{equation}{section}
\setcaptionmargin{1cm}

\def\La{\mathcal{L}}

\def\({\left(}
\def\){\right)}
\def\f{\frac}
\def\be{\begin{equation}}
\def\ee{\end{equation}}

\def\de{\partial}
\def\demub{\de_{\mu}}
\def\denub{\de_{\nu}}

\def\demua{\de^{\mu}}

\begin{document}
\begin{titlepage}
%\begin{flushright}
%\end{flushright}
\vskip 1.0cm
\begin{center}
%Vector pair production at the LHC by gluon fusion in an Effective Lagrangian approach
{\Large \bf Top quark effects in composite vector pair production at the LHC} \vskip 1.0cm
{\large A. E. C\'arcamo Hern\'andez$^{a}$}\\[0.7cm]
{\it $^a$ Universidad T\'ecnica Federico Santa Mar\'{\i}a and Centro Cient\'{\i}fico-Tecnol\'ogico de Valpara\'{\i}so, Casilla 110-V, Valpara\'{\i}so, Chile}
\\[5mm]
{antonio.carcamo@usm.cl}
\end{center}
\vskip 1.0cm
\begin{abstract}
In the context of a strongly coupled Electroweak Symmetry Breaking, composite light scalar singlet and composite triplet of heavy vectors may arise from an unspecified strong dynamics and the interactions among themselves and with the Standard Model gauge bosons and fermions can be described by a $SU(2)_L\times SU(2)_R/SU(2)_{L+R}$ Effective Chiral Lagrangian. In this framework, the production of the $V^{+}V^{-}$ and $V^{0}V^{0}$ final states at the LHC by gluon fusion mechanism is studied in the region of parameter space consistent with the unitarity constraints in the elastic channel of longitudinal gauge boson scattering and in the inelastic scattering of two longitudinal Standard Model gauge bosons into Standard Model fermions pairs. The expected rates of same-sign di-lepton and tri-lepton events from the decay of the $V^{0}V^{0}$ final state are computed and their corresponding backgrounds are estimated. It is of remarkable relevance that the $V^{0}V^{0}$ final state can only be produced at the LHC via gluon fusion mechanism since this state is absent in the Drell-Yan process. It is also found that the $V^{+}V^{-}$ final state production cross section via gluon fusion mechanism is comparable with the $V^{+}V^{-}$ Drell-Yan production cross section. The comparison of the $V^{0}V^{0}$ and $V^{+}V^{-}$ total cross sections will be crucial for distinguishing the different models since the vector pair production is sensitive to many couplings. This will also be useful to determine if the heavy vectors are only composite vectors or are gauge vectors of a spontaneously broken gauge symmetry.  
\end{abstract}
\vskip 1cm \hspace{0.7cm}
\end{titlepage}

%\tableofcontents

\section{Introduction.}
The ATLAS and CMS experiments at the CERN Large Hadron Collider (LHC) have found a $125$ GeV Higgs boson, increasing our knowledge of the Electroweak Symmetry Breaking (EWSB) sector and opening a new era in particle physics. It remains to study whether the new observed scalar state comes from a weakly or strongly coupled dynamics responsible for EWSB. A weakly coupled dynamics describing the mechanism of EWSB is provided by the Standard Model and its Supersymmetric extensions. Now the priority of the LHC experiments will be to measure precisely the couplings of the new particle to fermions and gauge bosons and to establish its quantum numbers in order to determine if the recently discovered Higgs boson is a weakly or a strongly coupled state. It also remains to look for further new states associated with the EWSB mechanism which will allow to discriminate among the different theoretical models addressed to explain EWSB.\newline
In spite of the very good agreement of the Standard Model predictions with experimental data, the Standard Model has the hierarchy problem, which is the instability of the mass of the Higgs field against quantum corrections, which are proportional to the square of the cut-off. This means that in a quantum theory with a cut-off at the Planck scale $\Lambda\simeq 10^{19}$ GeV, the Higgs boson mass will have quantum corrections that will raise it to about the Planck scale, unless an extreme fine-tuning of $34$ decimals is performed in the bare mass. This is the naturalness problem of the Standard Model. \newline
This problem can be overcome if one considers EWSB mechanisms in the framework of strongly interacting dynamics, where the theory becomes non-perturbative above the Fermi scale, and where the breaking is achieved through some condensate \cite{Appelquist:2003,Hirn:2007,Belyaev:2008yj,Quigg:2009,Hill:2003,Andersen:2011}. In the strongly interacting picture of EWSB, many models have been proposed which predict the existence of composite particles, e.g. composite scalars \cite{Kaplan:1983,Luty:1990, Chivukula:1993, Contino:2009,Contino:2010t, Grojean, ggpr, Burdman:2007, Low:2009, Zerwekh:2010,Burdman:2011,Carcamo:2012}, composite vector resonances \cite{Bagger,Pelaez:1996,SekharChivukula:2001, Csaki:2003, Barbieri:2008,Cata:2009iy,Barbieri:2010,Torre:2011a}, composite scalar and vector resonances \cite{Casalbuoni:1985,Zerwekh:2006,Carcamo:2010,Carcamo:2011,Torre:2011b} and composite fermions \cite{Kaplan:1991dc,Barbieri:2008b}. The spin-0 and spin-1 resonances predicted by these models play a very important role in controlling unitarity in longitudinal gauge boson scattering up to the cut-off $\Lambda \simeq 4\pi v$ \cite{Chivukula:2003,Barbieri:2003pr,Nomura2003,Foadi:2003xa,Georgi:2004iy,SekharChivukula:2008mj,Foadi:2008,Sanino:2008,Barbieri:2010}. For appropiate couplings and masses, the composite resonances may also account for the Electroweak Precision Tests. Furthermore, a composite scalar does not have the hierarchy problem since quantum corrections to its mass are saturated at the compositeness scale.\newline
The phenomenology of heavy vector states at high-energy colliders~\cite{He:2007ge,Accomando:2008jh,Birkedal:2005yg,Han,Belyaev:2008yj}, as well as their role in electroweak observables, is subject of intensive discussion. However, in most of the existing analyses specific dynamical assumptions are made such as considering  these vector states as the gauge vectors of a spontaneously broken gauge symmetry. Recent studies~\cite{Hirn:2007,Barbieri:2008} show that these assumptions may be too restrictive for generic models based on strong dynamics at the TeV scale, and only going beyond these assumptions, can one successfully account for the EWPT, by considering only exchange of heavy vectors. \newline
In the most general framework of strongly interacting dynamics for Electroweak Symmetry Breaking, one can have composite resonances which could be spin $0$, $1/2$ and $1$ states as well as tensor resonances or resonances of even higher spin. These composite particles are bound states of more fundamental constituents which are held together by a new strong interaction and their discovery could provide the first clue of strong EWSB at the LHC. If not too heavy, say below 1 TeV, the single production, either by Vector Boson Fusion (VBF) or by the Drell--Yan (DY) process, or its production in association with a standard gauge boson are very likely to be the first manifestations of $V$ at the LHC \cite{Belyaev:2008yj,Cata:2009iy,He:2007ge,Accomando:2008jh,Birkedal:2005yg}. To understand the underlying dynamics, several measurements and observations will certainly be required.\newline
It is assumed that this new strong dynamics supposedly breaking the Electroweak Symmetry is by itself invariant under a global $SU(2)_{L}\times SU(2)_{R}$ symmetry, which is spontaneously broken to the diagonal $SU(2)_{L+R}$ subgroup. After gauging the Standard Model gauge group, the $SU(2)_L\times SU(2)_R$ global symmetry of the new strong dynamics is broken down to the $SU(2)_{L+R}$ custodial group. It is also assumed that the strong dynamics responsible for EWSB, gives rise to a composite triplet of heavy vectors $V^a$ degenerate in mass belonging to the adjoint representation of the $SU(2)_{L+R}$ custodial symmetry group and to a composite scalar singlet $h$. The Lagrangian that describe this model, for some specific choice of the parameters can be obtained from a gauge theory based on $SU\(2\)_{L}\times SU\(2\)_{C}\times U\(1\)_{Y}$ spontaneously broken by two Higgs doublets (with the same \textsc{vev}) in the limit $m_{H}\gg \Lambda$ for the mass of the $L$-$R$-parity odd scalar $H$ \cite{Carcamo:2010}. In this framework, the role of unitarization of the different scattering channels is played both by the scalar and the vector (an example of this phenomenon is discussed for technicolor models in \cite{Foadi:2008}. This setup could possibly explain the excess of events in the $h\to\gamma\gamma$ decay recently observed at the LHC, since the heavy vectors in the triangular loop give a contribution to this process. Furthermore, I do not consider states of spin $2$ and higher since they are in general heavier than the states of lower spin that would first be discovered at the LHC.\newline
In the aforementioned framework, previously studied in \cite{Carcamo:2010}, I introduce in Section 2 the relevant $SU(2)_L\times SU(2)_R/SU(2)_{L+R}$ effective chiral Lagrangian which describes the composite scalar singlet and the composite triplet of heavy vectors with masses below the cut-off $\Lambda\simeq 4\pi v\approx 3$ TeV, the interactions among themselves and with the SM gauge bosons and SM fermions. Since, in this general framework, the SM fermions have (proto)-Yukawa interactions with the light composite scalar, since this scalar interacts with a heavy composite vector pair, and considering the large rate of gluons at the LHC, the top quark effects will be relevant for the vector pair production at the LHC via a gluon fusion mechanism, through a triangular loop followed by a scalar propagator. That is why, in this context, the vector pair production via the gluon fusion mechanism can compete with the vector pair production via Drell-Yan annihilation discussed in \cite{Barbieri:2010}. It is of particular relevance the presence of the $V^{0}V^{0}$ final state in the gluon fusion mechanism, which is absent in the Drell-Yan process. \newline
The absence of the $V^{0}V^{0}$ final state in the Drell-Yan process is due to the fact that the $ZV^{0}V^{0}$ and $V^{0}V^{0}V^{0}$ couplings do not exist, they are forbidden by gauge invariance. The absence of such couplings can be seen from the first term of expression (\ref{eq7}) and from the third and fourth terms of expression (\ref{LVint}). Figure 1 shows the Feynman diagrams corresponding to the Drell-Yan production of a single and a pair of heavy vectors. The Drell-Yan production of a single heavy vector has been previously studied and the corresponding total cross section for a charged vector resonance has been found to be between $0.2$ pb-$11$ pb \cite{Cata:2009iy}. That is why I am not discussing any of the Drell-Yan single production processes of the heavy vectors.
 %The dominant mechanism for producing a single heavy vector state is the Drell-Yan production process which has been previosly discussed \cite{Barbieri:2008}.
 %The single heavy vector production mechanism via vector boson fusion has been previously discussed \cite{Barbieri:2008}. In particular, the total cross section for the process $pp\to V+2jets\to WZ+2jets\to 3leptons+2jets+{\displaystyle{\not }}E_{T}$ at the LHC has been found to be between $0.3$fb and $3$fb \cite{Barbieri:2008}. 
In Section 3, the squared amplitudes for the composite heavy vector pair production via the gluon fusion mechanism summed over the polarization and color states are computed. In Section 4, the total cross sections for the production of the $V^{+}V^{-}$ and $V^{0}V^{0}$ final states at the LHC are computed for different values of the parameters consistent with the unitarity constraints in the elastic channel of longitudinal gauge boson scattering and in the inelastic scattering of two longitudinal Standard Model gauge bosons into Standard Model fermions pairs \footnote{I shall not impose the constraints coming from the EW Precision Tests since further effects can be present, e.g. due to new fermionic degrees of freedom, that obscure their interpretation and/or a strong sensitivity to the physics at the cut-off may be involved which I do not pretend to control.}. The discussion of the phenomenology of the same-sign di-lepton and tri-lepton events at the LHC for an integrated luminosity of $100$ fb$^{-1}$ and a comparison of these events with the backgrounds are presented in Section 5. The conclusions are given in Section 6.
\begin{figure}[tbh]
\includegraphics[width=8cm,height=7cm,angle=0]{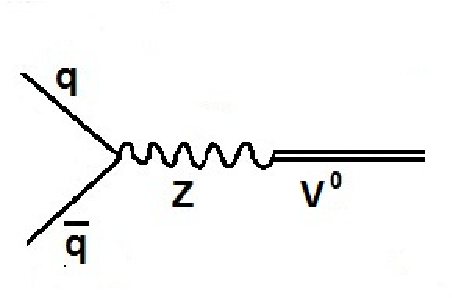}\includegraphics[width=8cm,height=7cm,angle=0]{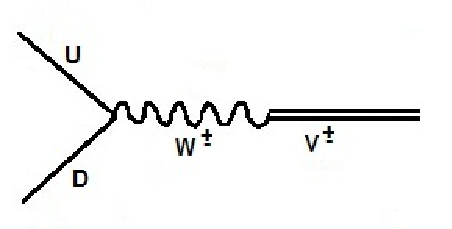}
\includegraphics[width=8cm,height=7cm,angle=0]{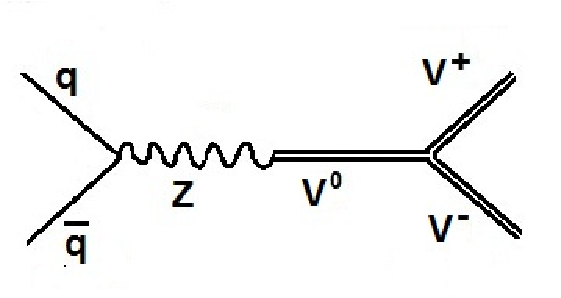}\includegraphics[width=8cm,height=7cm,angle=0]{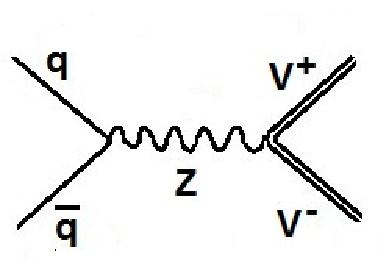}
\caption{Feynman diagrams corresponding to the Drell Yan production of a single and a pair of heavy vectors.}
\label{fig1b}
\end{figure}
\section{Chiral Lagrangian with massive spin one fields, scalar singlet and SM fermions.}
The starting point is the usual lowest order chiral Lagrangian for the $SU(2)_L\times SU(2)_R/SU(2)_{L+R}$ Goldstone fields with the addition of the
invariant kinetic terms for the $W$ and $B$ bosons \cite{Barbieri:2010}:
\be\label{lbasic}
\La_{\chi}=\f{v^{2}}{4}\left<D_{\mu}U\(D^{\mu}U\)^{\dag}\right>-\f{1}{2g^{2}}\left<W_{\mu\nu}W^{\mu\nu}\right>-\f{1}{2g^{\prime 2}}\left<B_{\mu\nu}B^{\mu\nu}\right>\,,
\ee
where
\be
\begin{array}{ll}
\displaystyle D_{\mu}U=\demub U-iB_{\mu}U+iU W_{\mu}\,,\qquad  & U=e^{\f{i\pi}{v}}\,, \qquad \pi=\pi^{a}\tau^{a}\,,\\
\displaystyle W_{\mu\nu}=\demub W_{\nu}-\denub W_{\mu}-i[W_{\mu},W_{\nu}]\,,\qquad & W_{\mu}=\f{g}{2}W_{\mu}^{a}\tau^{a}\,,\\
\displaystyle B_{\mu\nu}=\demub B_{\nu}-\denub B_{\mu}\,, & B_{\mu}=\f{g'}{2}B_{\mu}^{0}\tau^{3}\,,\\
\end{array}
\ee 
$U$ being the matrix which contains the Goldstone boson fields $\pi^a$ with $a=1,2,3$, the $\tau^{a}$ are the ordinary Pauli matrices and $\left\langle{}\right\rangle$ denotes the trace over $SU(2)$.\newline 
Now, a heavy spin-1 state belonging to the adjoint representation of $SU(2)_{L+R}$ is considered, so that $V_{\mu}=\f{1}{\sqrt{2}}V_{\mu}^{a}\tau^{a}$. The $SU(2)_L\times SU(2)_R$-invariant kinetic Lagrangian for the heavy
spin-1 fields is given by 
\begin{equation} 
\mathcal{L}_{\text{kin}}^V=-\frac{1}{4}\left\langle \hat{V}^{\mu\nu}\hat{V}%
_{\mu\nu}\right\rangle +\frac{M_{V}^{2}}{2}\left\langle {V}^{\mu}{V}%
_{\mu}\right\rangle \,.
\label{eq7}
\end{equation}
The field strength tensor $\hat{V}_{\mu\nu}=\nabla_{\mu}V_{\nu}-\nabla_{\nu}V_{\mu}$ is written in terms of the $SU(2)_{L}\times SU(2)_{R}$ covariant derivative
\be
\nabla_{\mu}V_{\nu}=\demub V_{\nu}+[\Gamma_{\mu},V_{\nu}]\,,
\ee
with the connection $\Gamma_{\mu}$ given by
\be
\Gamma_{\mu}=\f{1}{2}\Big[u^{\dag}\(\demub-iB_{\mu}\)u+u\(\demub-iW_{\mu}\)u^{\dag}\Big]\,,\qquad u\equiv \sqrt{U}\,,\qquad \Gamma_{\mu}^{\dag}=-\Gamma_{\mu}\,.
\ee
Assuming that the new strong dynamics is invariant under parity and considering the heavy vector states as the gauge vectors of a spontaneously broken symmetry, the interaction Lagrangian of the heavy vector with the SM model gauge fields and with the Goldstone bosons has been found to be given by \cite{Barbieri:2010}
\be
\begin{array}{lll}
\displaystyle \La^{V}_{\text{int}}&=&\displaystyle -\f{ig_{V}}{2\sqrt{2}}\left<\hat{V}_{\mu\nu}[u^{\mu},u^{\nu}]\right>-\f{g_{V}}{\sqrt{2}}\left<\hat{V}_{\mu\nu}\(uW^{\mu\nu}u^{\dag}+u^{\dag}B^{\mu\nu}u\)\right>+\frac{i}{2}\left\langle V_{\mu }V_{\nu }\left( uW^{\mu \nu }u^{\dagger }+u^{\dagger}B^{\mu \nu }u\right) \right\rangle\\\\
&&\displaystyle +\f{ig_{K}}{4\sqrt{2}}\left<\hat{V}_{\mu\nu}[V^{\mu},V^{\nu}]\right>-\f{1}{8}\left<[V_{\mu},V_{\nu}][u^{\mu},u^{\nu}]\right>+\f{g_{V}^{2}}{8}\left<[u_{\mu},u_{\nu}][u^{\mu},u^{\nu}]\right>\,,
\label{LVint}
\end{array}
\ee 
where $u_{\mu}=u^{\dag}_{\mu}=iu^{\dag}D_{\mu}U u^{\dag}$.\newline
A composite scalar singlet which could be a Strongly Interacting Light Higgs (SILH) boson in the sense of \cite{ggpr} or a more complicated object arising from an unknown strong dynamics is also considered. The Lagrangian which includes the kinetic and mass terms for the scalar as well as the interactions of this scalar with the Goldstone bosons, SM gauge fields and SM fermions is given by \cite{Contino:2010}
\be\label{lscalar}
\La_{h}=\f{1}{2}\demub h \demua h +\f{m_{h}^{2}}{2}h^{2}+\f{v^{2}}{4}\left<D_{\mu}U\(D^{\mu}U\)^{\dag}\right>\(2a\f{h}{v}+b\f{h^{2}}{v^{2}}\)- \frac{v}{\sqrt{2}} \sum_{i,j} \left( \bar u_L^{(i)} d_L^{(i)} \right) U \left( 1+ c\, \frac{h}{v}\right)\begin{pmatrix} \lambda_{ij}^u\,  u_R^{(j)} \\[0.1cm] \lambda_{ij}^d\,  d_R^{(j)} \end{pmatrix} + h.c.
\ee
where $\lambda_{ij}^u$ and $\lambda_{ij}^d$ are the up and down type quarks Yukawa couplings, respectively.\newline
The Lagrangian $\La_{h-V}$ which describes the interaction between the scalar and the heavy vector $V$ is \cite{Carcamo:2010}
\be\label{lvhinteraction}
\La_{h-V}=\f{dv}{8g_{V}^{2}}h\left<V_{\mu}V^{\mu}\right>.
\ee
Here $a$, $b$, $c$, $d$, $g_V$ and $g_K$ are dimensionless constants~\footnote{In general $c$ will be a matrix in flavor space, but in the following it is assumed for simplicity that it is proportional to unity in the basis in which the mass matrix is diagonal. This guarantees the absence of flavor changing neutral effects originated from the tree level exchange of $h$.}.\newline
%where $c$ is a dimensionless constant
Summarizing, in the framework of strongly interacting dynamics for EWSB, the interactions among the composite scalar singlet, composite triplet of heavy vectors and the SM gauge bosons and SM fermions can be described by the following model independent $SU(2)_L\times SU(2)_R/SU(2)_{L+R}$ chiral Lagrangian:
\be\label{lvector}
\La_{\text{eff}}=\La_{\chi}+\La^{V}_{\text{kin}}+\La^{V}_{\text{int}}+\La_{h}+\La_{h-V}.
\ee
%This Lagrangian is based in the following assumptions:
Here the following assumptions have been made:
\begin{enumerate}
\item Before weak gauging, the Lagrangian responsible for EWSB has a $SU(2) _{L}\times SU(2) _{R}$ global symmetry which is spontaneously broken by the new strong dynamics down to the $SU(2)_{L+R}$ custodial group. The spontaneous breaking of the global symmetry also leads to the breaking of the standard electroweak gauge symmetry, $SU(2)_L\times U(1)_Y$, down to the electromagnetic $U(1)$.
\item The strong dynamics produces a composite triplet of heavy vectors degenerate in mass belonging to the $SU(2)_{L+R}$ adjoint representation and a composite scalar singlet under $SU(2)_{L+R}$. 
\item Only one vector triplet $V_{\mu}^{a}$ of the  $SU\left(2\right)_{L+R}$ group has a mass below the cut-off $\Lambda\approx 3\,\text{TeV}$, while the parity odd heavy vectors are integrated out since they are assumed to be heavier than the heavy vectors. The new $V$ states couple to fermions only via SM gauge interactions.
\item The light scalar singlet of mass $m_{h}=125$ GeV interacts with the Standard Model gauge bosons and fermions only via weak gauging and (proto)-Yukawa couplings, respectively.
\end{enumerate}
In the model under consideration, the interactions among the heavy vector states and with the Standard Model gauge fields and Goldstone bosons have been discussed in Ref.\cite{Barbieri:2010}, the interaction between the scalar and the heavy vector $V$ has been studied in Ref.\cite{Carcamo:2010} while the interactions between the composite scalar and the SM particles have been introduced in Ref.\cite{Contino:2010}.\newline
It is worth to mention that the custodial symmetry $SU(2)_{L+R}$ keeps the heavy vectors at the same mass at tree level; however, the operator involving $2V$'s with one $B$ (third term in expression (\ref{LVint})) generates a splitting between the neutral and charged heavy vector masses at one loop level. Besides that, the effective $SU(2)_L\times SU(2)_R/SU(2)_{L+R}$ chiral Lagrangian given in (\ref{lvector}) is invariant under parity at tree level, but at one loop level the breaking of parity takes place in an analogous way as the anomalous $U(1)_A$ breaking in QCD. This breaking of parity results in different masses for the vectors and axial vectors.
In analogy with QCD, the fact that the heavy vectors (analogous to the $\rho$ meson) are lighter than the heavy axial vectors (analogous to the $a_1$ meson) is due in part to the higher orbital angular momentum and spin of the heavy axial vectors. \newline
\section{Gluon Fusion production amplitudes}
\begin{figure}[tbh]
\includegraphics[width=12cm,height=7cm,angle=0]{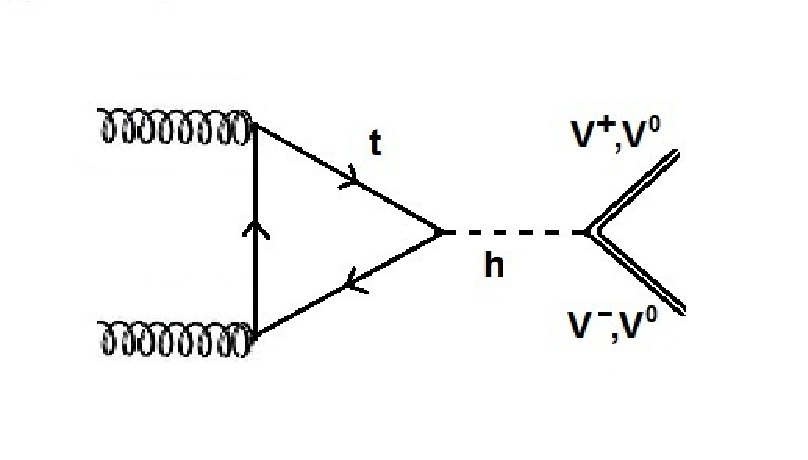}
%\centering[width=10cm,height=10cm,angle=0]
\caption{Leading order diagram, containing a $hVV$ coupling vertex, of the vector pair production through the gluon fusion process. Crossing the gluon legs yields a second diagram.}
\label{fig1a}
\end{figure}

There are several one loop level contributions to the amplitudes for the heavy vector pair production by the gluon fusion mechanism. They belong to two types, triangular and box diagrams containing a top quark running in them; they are shown in Figures \ref{fig1a}, \ref{fig1b} and \ref{fig1c}, respectively. The triangular loop can be followed by a scalar propagator, a $Z$ boson propagator, $Z$ and $V^0$ propagators coupled by the $Z$-$V^0$ mixing, $\pi^0$ and $Z$ propagators coupled by the $\pi^0$-$Z$ mixing. These possibilities define several one loop level contributions to the $gg\rightarrow V^{+}V^{-}$ scattering amplitude, proportional to $\frac{\alpha_S}{g^2_V}$, $g^2\alpha_S$, $g^2g_Kg_V\alpha_S$ and $g^2g_V\alpha_S$, respectively. This implies that the only relevant contribution to the $gg\rightarrow V^{+}V^{-}$ scattering amplitude is the one having the triangular loop followed by a scalar propagator coupled to it. At one loop level, the only top quark triangular diagram contribution to the $gg\rightarrow V^{0}V^{0}$ scattering amplitude, is the one containing a scalar propagator coupled to the triangular loop and is shown in Figure \ref{fig1a}.\newline
\begin{figure}[tbh]
\includegraphics[width=5.5cm,height=7cm,angle=0]{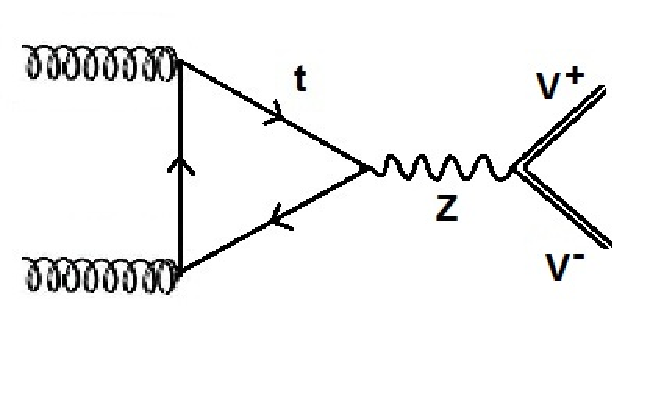}\includegraphics[width=5.5cm,height=7cm,angle=0]{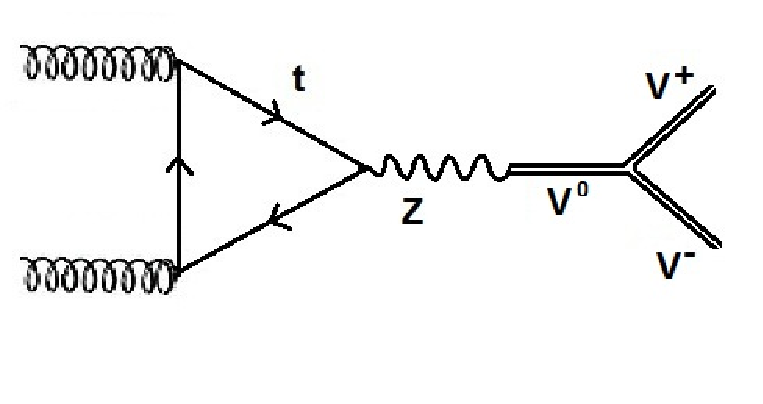}\includegraphics[width=5.5cm,height=6.5cm,angle=0]{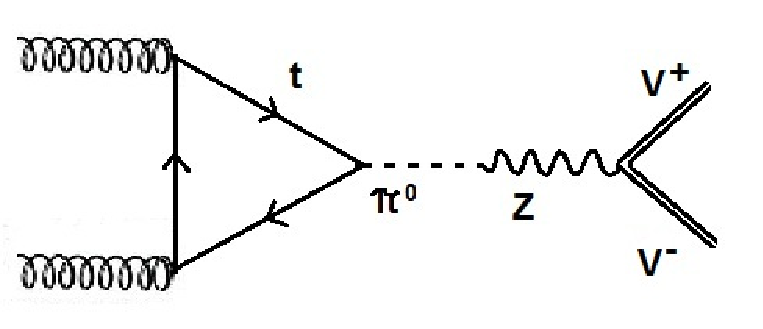}
\caption{Subleading order triangular diagrams of the vector pair production through the gluon fusion process.}
\label{fig1b}
\end{figure}
\begin{figure}[tbh]
\includegraphics[width=8cm,height=7cm,angle=0]{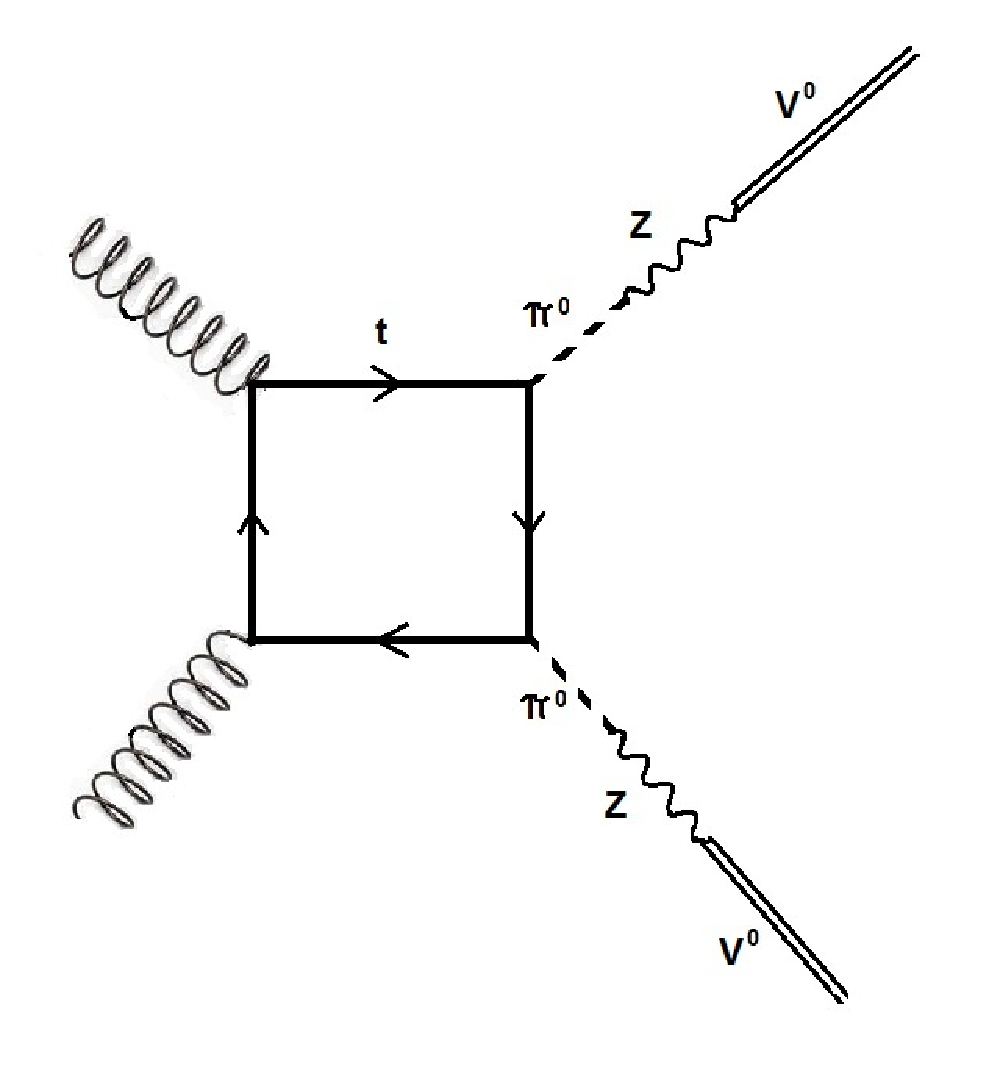}\includegraphics[width=8cm,height=7cm,angle=0]{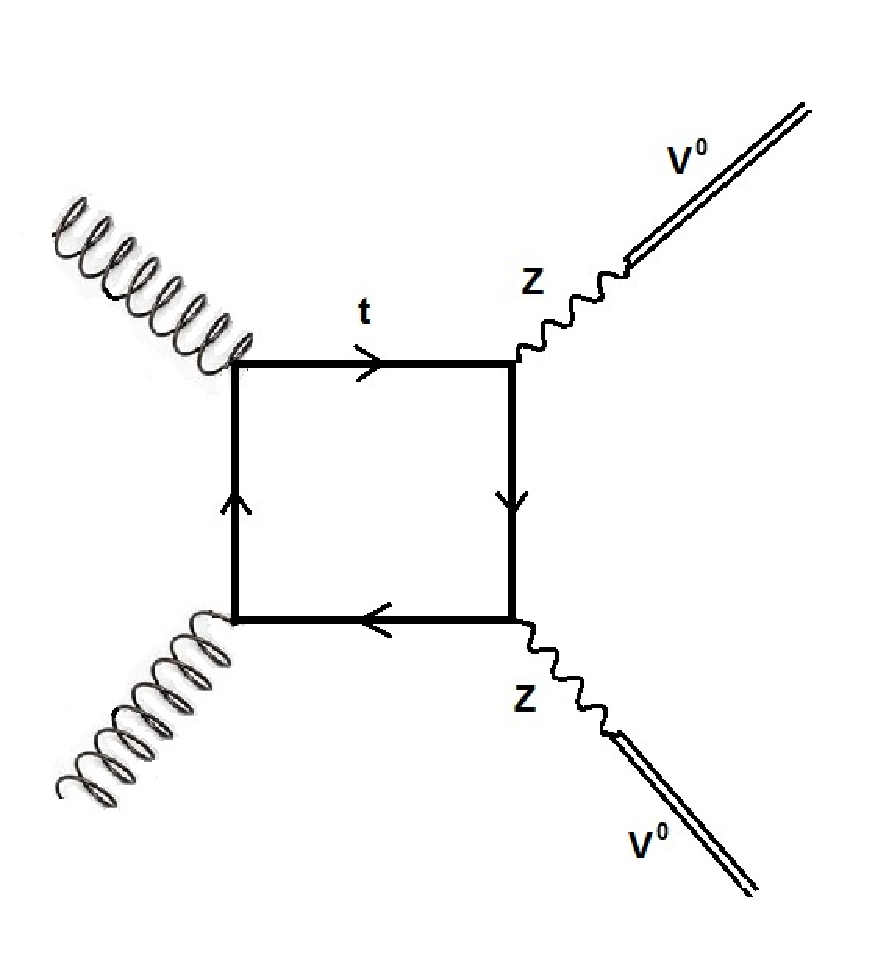}
%\centering[width=10cm,height=10cm,angle=0]
%\input{Diag1.ps} 
\caption{Subleading order box diagrams of the vector pair production through the gluon fusion process.}
\label{fig1c}
\end{figure}
The one loop level box diagrams only contribute to the $gg\rightarrow V^{0}V^{0}$ scattering amplitude since they can be followed by a $Z$ boson propagator, $\pi^0$ and $Z$ propagators coupled by the $\pi^0$-$Z$ mixing. This implies that at one loop level there are only two box diagram contributions to the $gg\rightarrow V^{0}V^{0}$ scattering amplitude, both of them proportional to $g^4g^2_V\alpha_S$, so that they can be neglected. Then, the only relevant contributions to the amplitudes, for the heavy vector pair production by the gluon fusion mechanism, come from the top quark in the triangular loop followed by a scalar propagator. This implies that the amplitude for the gluon fusion process $gg\rightarrow V^{+}V^{-}$ is given by the following expression:
\begin{eqnarray}
A\left( gg\rightarrow V^{+}V^{-}\right) &=&-\frac{\alpha _{S}}{\pi}\left( 
\frac{cd}{8g_{V}^{2}\left( s-M_{h}^{2}\right) }\right) \delta _{ab}\left[
g^{\mu \nu }\left( p\cdot k\right) -p^{\nu }k^{\mu }\right]I\left( \frac{s}{%
m^{2}_t}\right) \varepsilon _{\mu }\left( p,\chi \right) \varepsilon _{\nu
}\left( k,\chi ^{\prime }\right)  \notag \\
&&\times g^{\rho \sigma }\varepsilon _{\rho }\left( l,\xi \right)
\varepsilon _{\sigma }\left( q,\xi ^{\prime }\right), \label{g2}
\end{eqnarray}
where $I\left( \frac{s}{m^{2}_t}\right) $ is given by
\begin{equation}
I\left( \frac{s}{m^{2}_t}\right) =\int_{0}^{1}dx\int_{0}^{1-x}dy\frac{1-4xy}{1-%
\frac{s}{m^{2}_t}xy}.  \label{g1a}
\end{equation}
Here $m_t$ is the mass of the top quark, $%
\varepsilon _{\mu }\left( p,\chi \right) $ and $\varepsilon _{\nu }\left(
k,\chi ^{\prime }\right) $ are the polarization vectors of the gluons,  $%
\varepsilon _{\mu }\left( l,\xi \right) $ and $\varepsilon _{\nu }\left(
q,\xi ^{\prime }\right) $ are the polarization vectors of the heavy vectors, $%
s=\left( p+k\right) ^{2}=2p\cdot k$ is the energy of the virtual scalar, $a,b=1,2,\ldots ,8$ are the color indices of the gluons. Moreover a factor of $2$ has been included in the expression (\ref{g2}) to take into account the diagram where the gluon legs are crossed.\newline
Besides that, in order to cancel the growth of the $\pi \pi \rightarrow \bar{\psi}\psi$ scattering amplitude with $\sqrt{s}$ (where $\psi$ denotes a SM fermion in the mass eigenstate), $c$ should satisfy
\begin{equation}
c=\frac{1}{a}.
\end{equation}
It is shown in Ref.\cite{Carcamo:2010} that the elastic $W_LW_L$ scattering amplitude has a good asymptotic behavior provided that
\begin{equation}
a=\sqrt{1-\frac{3G_{V}^{2}}{v^{2}}},\hspace{1.7cm}\hspace{1.7cm}%
G_{V}=g_{V}M_{V}, \label{ga1}
\end{equation}
which implies the upper bound  $G_{V}\leq v/\sqrt{3}$ for the coupling $G_V$ of the heavy vector to two longitudinal SM gauge bosons.\newline
The previous conditions imply that the coupling $c$ should satisfy the following relation:
\begin{equation}
c=\frac1{\sqrt{1-\frac{3G_{V}^{2}}{v^{2}}}}.\label{cc}
\end{equation}
From expression (\ref{g2}) and taking into account that the symmetry factor of the $hV^{0}V^{0}$ vertex is $2$, it follows that the squared amplitudes for the vector pair production via the gluon fusion mechanism summed over the polarization and color states are given by
%over the color indices of the gluons and over the initial and final state polarizations is given by:
%By performing the sum over the color indices of the gluons and over the initial and final state polarizations in the square of the amplitude for the gluon fusion process $gg\rightarrow V^{+}V^{-}$ with a heavy fermion in a triangular loop and a virtual higgs which decays in heavy vector pairs, the following expression is obtained:
\begin{eqnarray}
\sum_{a,b,\chi ,\chi ^{\prime },\xi ,\xi ^{\prime }}\left\vert A\left(
gg\rightarrow V^{+}V^{-}\right) \right\vert ^{2} &=&\frac{1}{4}%
\sum_{a,b,\chi ,\chi ^{\prime },\xi ,\xi ^{\prime }}\left\vert A\left(
gg\rightarrow V^{0}V^{0}\right) \right\vert ^{2}  \notag \\
&=&\frac{c^{2}d^{2}\alpha _{S}^{2}s^{2}}{16\pi ^{2}g_{V}^{4}\left(
s-M_{h}^{2}\right) ^{2}}\left\vert I\left( \frac{s}{m^{2}_t}\right)
\right\vert ^{2}\left( \frac{s^{2}}{4M_{V}^{4}}-\frac{s}{M_{V}^{2}}+3\right).
\label{rr}
\end{eqnarray}
The gluon fusion vector pair production amplitudes grow as $\frac{s}{M_{V}^{2}}$ at high energies. In this case the asymptotic behavior of the gluon fusion vector pair production amplitudes will have to be improved by introducing a scalar-vector mixing term, with appropiate coupling.
\section{Vector pair production total cross sections via gluon fusion}
\begin{figure}
% GNUPLOT: LaTeX picture with Postscript
\begingroup%
\makeatletter%
\newcommand{\GNUPLOTspecial}{%
  \@sanitize\catcode`\%=14\relax\special}%
\setlength{\unitlength}{0.0500bp}%
\begin{picture}(8640,5040)(0,0)%
  \special{psfile=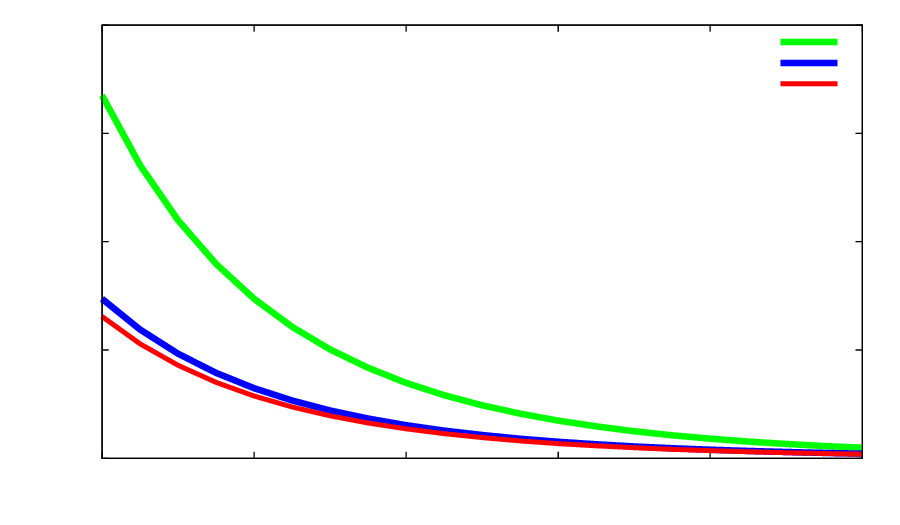 llx=0 lly=0 urx=432 ury=252 rwi=4320}
  \put(7375,4236){\makebox(0,0)[r]{\strut{}$\sigma\left(pp\rightarrow gg\rightarrow V^{+}V^{-}\right)$ for $G_V=v/2$ and $d=1$}}%
  \put(7375,4436){\makebox(0,0)[r]{\strut{}$\sigma\left(pp\rightarrow gg\rightarrow V^{+}V^{-}\right)$ for $G_V=v/\sqrt{6}$ and $d=1$}}%
  \put(7375,4636){\makebox(0,0)[r]{\strut{}$\sigma\left(pp\rightarrow gg\rightarrow V^{+}V^{-}\right)$ for $G_V=\sqrt{5}v/4$ and $d=1$}}%
  \put(4629,140){\makebox(0,0){\strut{}$M_V$(GeV)}}%
  \put(160,2719){%
  \special{ps: gsave currentpoint currentpoint translate
630 rotate neg exch neg exch translate}%
  \makebox(0,0){\strut{}$\sigma (fb)$}%
  \special{ps: currentpoint grestore moveto}%
  }%
  \put(8278,440){\makebox(0,0){\strut{} 1000}}%
  \put(6818,440){\makebox(0,0){\strut{} 900}}%
  \put(5359,440){\makebox(0,0){\strut{} 800}}%
  \put(3899,440){\makebox(0,0){\strut{} 700}}%
  \put(2440,440){\makebox(0,0){\strut{} 600}}%
  \put(980,440){\makebox(0,0){\strut{} 500}}%
  \put(860,4799){\makebox(0,0)[r]{\strut{} 0.8}}%
  \put(860,3759){\makebox(0,0)[r]{\strut{} 0.6}}%
  \put(860,2720){\makebox(0,0)[r]{\strut{} 0.4}}%
  \put(860,1680){\makebox(0,0)[r]{\strut{} 0.2}}%
  \put(860,640){\makebox(0,0)[r]{\strut{} 0}}%
\end{picture}%
\endgroup
 
\vspace{-4mm}\caption{Total cross sections for the $V^{+}V^{-}$ production via the gluon fusion mechanism at the LHC for $\protect\sqrt{S}=14$ TeV and $d=1$ as functions of the heavy vector mass $M_V$ for different values of the $G_V$ parameter. The green, blue and red lines correspond to $G_V=\sqrt{5} v/4$, $G_V=v/\sqrt{6}$ and $G_V=v/2$ (gauge model scenario), respectively. Here $\protect\alpha _{S}=0.12$, $M_{h}=125$ GeV, $m_{t}=171.3$ GeV and $\mu=2M_V$. The coupling $c$ is chosen to satisfy the condition given in (\ref{cc}). Color Figure online.}
\label{fig1d}
% GNUPLOT: LaTeX picture with Postscript
\begingroup%
\makeatletter%
\newcommand{\GNUPLOTspecial}{%
  \@sanitize\catcode`\%=14\relax\special}%
\setlength{\unitlength}{0.0500bp}%
\begin{picture}(9000,5040)(0,0)%
  \special{psfile=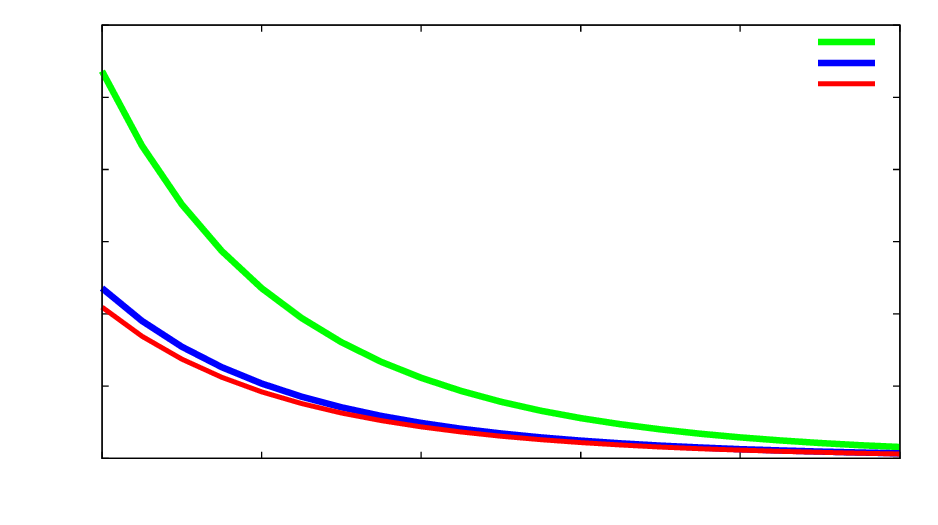 llx=0 lly=0 urx=450 ury=252 rwi=4500}
  \put(7735,4236){\makebox(0,0)[r]{\strut{}$\sigma\left(pp\rightarrow gg\rightarrow V^{0}V^{0}\right)$ for $G_V=v/2$ and $d=1$}}%
  \put(7735,4436){\makebox(0,0)[r]{\strut{}$\sigma\left(pp\rightarrow gg\rightarrow V^{0}V^{0}\right)$ for $G_V=v/\sqrt{6}$ and $d=1$}}%
  \put(7735,4636){\makebox(0,0)[r]{\strut{}$\sigma\left(pp\rightarrow gg\rightarrow V^{0}V^{0}\right)$ for $G_V=\sqrt{5}v/4$ and $d=1$}}%
  \put(4809,140){\makebox(0,0){\strut{}$M_V$(GeV)}}%
  \put(160,2719){%
  \special{ps: gsave currentpoint currentpoint translate
630 rotate neg exch neg exch translate}%
  \makebox(0,0){\strut{}$\sigma (fb)$}%
  \special{ps: currentpoint grestore moveto}%
  }%
  \put(8638,440){\makebox(0,0){\strut{} 1000}}%
  \put(7106,440){\makebox(0,0){\strut{} 900}}%
  \put(5575,440){\makebox(0,0){\strut{} 800}}%
  \put(4043,440){\makebox(0,0){\strut{} 700}}%
  \put(2512,440){\makebox(0,0){\strut{} 600}}%
  \put(980,440){\makebox(0,0){\strut{} 500}}%
  \put(860,4799){\makebox(0,0)[r]{\strut{} 3}}%
  \put(860,4106){\makebox(0,0)[r]{\strut{} 2.5}}%
  \put(860,3413){\makebox(0,0)[r]{\strut{} 2}}%
  \put(860,2720){\makebox(0,0)[r]{\strut{} 1.5}}%
  \put(860,2026){\makebox(0,0)[r]{\strut{} 1}}%
  \put(860,1333){\makebox(0,0)[r]{\strut{} 0.5}}%
  \put(860,640){\makebox(0,0)[r]{\strut{} 0}}%
\end{picture}%
\endgroup
 
\vspace{-4mm}\caption{Total cross sections for the $V^{0}V^{0}$ production via the gluon fusion mechanism at the LHC for $\protect\sqrt{S}=14$ TeV and $d=1$ as functions of the heavy vector mass $M_V$ for different values of the $G_V$ parameter. The green, blue and red lines correspond to $G_V=\sqrt{5} v/4$, $G_V=v/\sqrt{6}$ and $G_V=v/2$ (gauge model scenario), respectively. Here $\protect\alpha _{S}=0.12$, $M_{h}=125$ GeV, $m_{t}=171.3$ GeV and $\mu=2M_V$. The parameter $c$ is chosen to satisfy the condition given in (\ref{cc}). Color Figure online.}
\label{fig1e}
\end{figure}

The final states for the vector pair production via the gluon fusion mechanism obviously are the charge states $V^{+}V^{-}$ and $V^{0}V^{0}$. The total cross section for the $V^{+}V^{-}(V^{0}V^{0})$ production through the gluon fusion mechanism in proton proton collisions with center of mass energy $\sqrt{S}$ is given by
\begin{equation}
\sigma _{pp\rightarrow gg\rightarrow V^{+}V^{-}(V^{0}V^{0})}\left( S\right) =\int_{\sqrt{%
\frac{2M_{V}^{2}}{S}}}^{1}dx\int_{\sqrt{\frac{2M_{V}^{2}}{S}}%
}^{1}dyf_{p/g}\left( x,\mu ^{2}\right) f_{p/g}\left( y,\mu ^{2}\right)
\sigma _{gg\rightarrow V^{+}V^{-}(V^{0}V^{0})}\left( s\right),\label{p9}
\end{equation}
where $s=xyS$ is the partonic center of mass energy, $f_{p/g}\left( x,\mu ^{2}\right) $ and $f_{p/g}\left( y,\mu^{2}\right) $ are the distributions of gluons in the proton which carry momentum fractions $x$ and $y$ of the proton, respectively. \newline
Here, the choice $\mu =2M_V$ for the factorization scale is made motivated by the fact that in the $WW$ production in the SM at NLO the factorization scale is taken to be equal to $2M_W$ as done in Ref.\cite{Grazzini}. Besides that, $\sigma _{gg\rightarrow V^{+}V^{-}}\left(
s\right) $ is the parton level cross section for the process $gg\rightarrow V^{+}V^{-}$ given by
\begin{eqnarray}
\sigma _{gg\rightarrow V^{+}V^{-}}\left( s\right) &=&\frac1{4}\sigma _{gg\rightarrow V^{0}V^{0}}\left( s\right)=\frac{1}{4}\times \frac{%
1}{64}\times \frac{1}{16\pi s^{2}}\int_{t_{\min }}^{t_{\max }}\sum_{a,b,\chi
,\chi ^{\prime },\xi ,\xi ^{\prime }}\left\vert A\left( gg\rightarrow
V^{+}V^{-}\right) \right\vert ^{2}d\widehat{t},\label{p10}
\end{eqnarray}
being $t_{\min }$ and $t_{\max }$ given by: 
\begin{equation}
t_{\min }=-\left( \sqrt{\frac{s}{4}}+\sqrt{\frac{s}{4}-M_{V}^{2}}\right)
^{2},\hspace{1.7cm}\hspace{1.7cm}t_{\max }=-\left( \sqrt{\frac{s}{4}}-\sqrt{%
\frac{s}{4}-M_{V}^{2}}\right) ^{2}.\label{sa7}
\end{equation}
In the expression (\ref{p10}), the factor $\frac{1}{4}$ is due to the
average over the transverse polarization states of the gluons and the factor 
$\frac{1}{64}$ comes from the average over the color states of the gluons. \newline
Figures \ref{fig1d} and \ref{fig1e} show the total cross sections at the LHC for the $V^{+}V^{-}$ and $V^{0}V^{0}$ production via the gluon fusion mechanism as functions of the heavy vector mass $M_V$ and for different values of the $G_V$ parameter taking the scalar-vector coupling $d$ equal to $1$. In order to study how a deviation from the gauge model scenario affects the heavy vector pair production in the model under consideration, the values $G_V=v/\sqrt{6}$ and $G_V=\sqrt{5}v/4$ are chosen for the $G_V$ coupling as well as the value $G_V=v/2$ predicted by the gauge model. The gauge model scenario corresponds to the case where the composite heavy vector states are the gauge vectors of a spontaneously broken symmetry and is characterized by $G_V=\frac{v}{2}$, $a=\frac1{2}$ and $d=1$ \cite{Carcamo:2010}. A deviation from the couplings predicted by the gauge model scenario could be an indication of new degrees of freedom such as axial vector resonances. The heavy vector mass has been taken to range from $500$ GeV to $1$ TeV. Here the top quark mass has been taken to be equal to $m_{t}=171.3$ GeV and the light scalar mass is equal to $M_h=125$ GeV. The coupling $c$ has been chosen to satisfy the condition given in expression (\ref{cc}) which guarantees unitarity in the elastic channel for longitudinal gauge boson scattering and in the inelastic scattering of two longitudinal SM gauge bosons into SM fermions pairs.
The values of the total cross sections at the LHC for the production of the $V^{+}V^{-}$ and $V^{0}V^{0}$ final states by the gluon fusion mechanism with the top quark in the triangular loop as functions of the different parameters are listed in Table \ref{table2}. The values of the parameters in Table \ref{table2} account for the gauge model scenario as well as for moderate deviations from this scenario. For the computation of the total cross sections, the MSTW2008 LO gluon distribution function has been used.
\begin{table}[tbh]
\begin{minipage}[b]{8.2cm}
\centering
\begin{tabular}[c]{|c|c|c|c|c|}
		\hline
		$G_{V}$ &  $a$ &  $d$ & $V^+V^-$ (fb) & $V^0V^0$ (fb) \\
		\hline
		$\sqrt{5} v/4$  &$1/4$ & $0$ & $0$& $0$ \\
		\hline
		$\sqrt{5} v/4$  &$1/4$ & $1$ & $0.67$ & $2.68$ \\
		\hline
		$\sqrt{5} v/4$  &$1/4$ & $2$ & $2.68$ & $11.72$ \\
		\hline
		$v/2$  & $1/2$ & $0$ & $0$ & $0$ \\
		\hline
		$v/2$  & $1/2$ & $1$ & $0.26$ & $1.04$ \\
		\hline
		$v/2$  & $1/2$ & $2$ & $1.04$ & $4.16$ \\
		\hline
		$v/\sqrt{6}$  & $1/\sqrt{2}$ & $0$ & $0$  & $0$ \\
		\hline
		$v/\sqrt{6}$  & $1/\sqrt{2}$ & $1$ & $0.29$  & $1.16$ \\
		\hline
		$v/\sqrt{6}$  & $1/\sqrt{2}$ & $2$ & $1.16$  & $4.64$\\
		\hline
\end{tabular}
\\\vspace{4mm} \footnotesize{(\ref{table2}.a)}
 \end{minipage}
\ \hspace{2mm} \hspace{3mm} \ 
\begin{minipage}[b]{8.2cm}
\centering
\begin{tabular}[c]{|c|c|c|c|c|}
		\hline
		$G_{V}$ &  $a$ &  $d$ &  $V^+V^-$ (fb) & $V^0V^0$ (fb) \\
		\hline
		$\sqrt{5} v/4$  &$1/4$ & $0$ & $0$ & $0$ \\
		\hline
		$\sqrt{5} v/4$  &$1/4$ & $1$ & $0.02$ & $0.08$ \\
		\hline
		$\sqrt{5} v/4$  &$1/4$ & $2$ & $0.08$ & $0.32$ \\
		\hline
		$v/2$  & $1/2$ & $0$ & $0$ & $0$ \\
		\hline
		$v/2$  & $1/2$ & $1$ & $0.01$ & $0.04$ \\
		\hline
		$v/2$  & $1/2$ & $2$ & $0.04$ & $0.16$ \\
		\hline
		$v/\sqrt{6}$  & $1/\sqrt{2}$ & $0$ & $0$  & $0$ \\
		\hline
		$v/\sqrt{6}$  & $1/\sqrt{2}$ & $1$ & $0.01$  & $0.04$ \\
		\hline
		$v/\sqrt{6}$  & $1/\sqrt{2}$ & $2$ & $0.04$  & $0.16$ \\
		\hline
\end{tabular}
\\\vspace{4mm} \footnotesize{(\ref{table2}.b)}
 \end{minipage}
\vspace{-5mm}
\caption{Total cross sections for the production of the $V^{+}V^{-}$ and $%
V^{0}V^{0}$ final states by gluon fusion at the LHC for $\protect\sqrt{S}=14$ TeV as functions of the different
parameters for $M_{V}=500$ GeV (\protect\ref{table2}.a) and $M_{V}=1$ TeV (\protect\ref{table2}.b). Here $\protect\alpha _{S}=0.12$, $M_{h}=125$ GeV and $m_{t}=171.3$ GeV while the factorization scale is taken to be equal to $2M_V$. The parameter $a$ is chosen to satisfy the condition given in (\ref{ga1}). The gauge model scenario corresponds to the case $G_V=\frac{v}{2}$, $a=\frac1{2}$ and $d=1$.}
\label{table2}
\end{table}
\newpage
The total cross sections at the LHC for the production of the $V^{+}V^{-}$ and $V^{0}V^{0}$ final states via the gluon fusion mechanism take their minimum values for the gauge model scenario. This implies that the parameters of the gauge model scenario damp the high energy behavior of the vector pair production amplitudes via the gluon fusion mechanism. It can be seen that the total cross sections at the LHC for the vector pair production via the gluon fusion mechanism are small to give rise to a signal for a large region of the parameter space. It is worth to mention that the gluon fusion mechanism is the only process leading to the $V^{0}V^{0}$ final state that cannot be produced via Drell-Yan annihilation. Regarding the production of the $V^{+}V^{-}$ final state via the gluon fusion mechanism, its corresponding total cross section is comparable with the $V^{+}V^{-}$ Drell-Yan production cross section, which is independent on the $G_V$ coupling and is obtained in \cite{Barbieri:2010}.
%It is worth to mention that the neutral heavy vector pair production via gluon fusion mechanism is comparable with the production of neutral heavy vector pairs via Vector Boson Fusion discussed . $W_L-W_L-V$
It is also important to mention that a weak coupling $G_{V}$ of the heavy vector with two longitudinal SM gauge bosons and a strong coupling $d$ of the scalar with the heavy vector pairs favors larger cross sections for the vector pair production via a gluon fusion mechanism, since the corresponding squared amplitudes are proportional to $\frac{d^{2}}{G_{V}^{4}}$. This implies that deviations from $d=1$ will result in a strong increase of the vector pair production cross sections via the gluon fusion mechanism.

\section{Same-sign di-lepton and tri-lepton events}\label{sec6}

Since the composite vectors decay mainly into $WW$ or $WZ$, with branching ratio very close to 1, the final state obtained from the vector pair production via gluon fusion will have four SM gauge bosons. Considering only the $e$ and $\mu$ leptons coming from the $W$ decays, the Table \ref{BR}, which shows the cumulative branching ratios for at least two same-sign leptons and three leptons in the $V^0V^0$ charge configuration, is obtained.
\begin{table}[htb!]
\centering
\small\begin{tabular}[c]{|c|c|c|c|}
		\hline
		Decay Mode & di-leptons ($\%$)& tri-leptons ($\%$)\\
		\hline  
		$V^{0}V^{0}$$\to$ $W^{+}W^{-}W^{+}W^{-}$ & $8.9$ & $3.2$\\
		\hline  
\end{tabular}\caption{Dominant decay mode and cumulative branching ratios for the $V^{0}V^{0}$ charge configuration. For the same-sign di-lepton and tri-lepton branching rations only the $e$ and $\mu$ leptons coming from the $W$ decays are considered.}\label{BR}
\end{table}\newline
Using the values of the cumulative branching ratios given in Table \ref{BR} and reference integrated luminosity of $\int\mathcal{L}dt=100~\text{fb}^{-1}$ for the LHC, the total number of same-sign di-lepton and tri-lepton events is obtained and shown in Table \ref{events}.
\begin{table}[htb!]
%\small\begin{minipage}[b]{8.2cm}
\centering
\begin{tabular}[c]{|c|c|c|c|}
		\hline
		$G_{V}$ &  $a$ & di-leptons & tri-leptons \\
		\hline
		$\sqrt{5}v/4$  & $1/4$ & $24$ & $9$ \\
		\hline
		$v/2$  & $1/2$ & $9$ & $3$ \\
		\hline
		$v/\sqrt{6}$  & $1/\sqrt{2}$ & $11$  & $4$ \\
		\hline
\end{tabular}
\\\vspace{3mm} %\footnotesize{(\ref*{events}.a)}
% \end{minipage}
\ \hspace{2mm} \hspace{3mm} \ 
\caption{Total number of same-sign di-lepton and tri-lepton events ($e$ or $\mu$ from $W$ decays) for the vector pair production via gluon fusion at the LHC for $\sqrt{S}=14$ TeV and $\int\mathcal{L}dt=100$ fb$^{-1}$ at $M_{V}=500$ GeV, $M_{h}=125$ GeV and $m_{t}=171.3$ GeV for different values of the parameter $G_{V}$ (or $a$ according to relation (\ref{ga1}) and for $d=1$. Since the gluon fusion total cross sections are proportional to $d^{2}$ the results can simply be generalized to different values of $d$. The choice $G_V=\frac{v}{2}$ and $a=\frac1{2}$ corresponds to the gauge model scenario.}\label{events}
\end{table}
% \newline
\begin{table}[htb!]
\centering
\small\begin{tabular}[c]{|c|c|}
\hline
\textbf{Signal}\protect & \textbf{Number of Events}\\
\hline
$pp\rightarrow V^{0}V^{0}\rightarrow W^{+}W^{-}W^{+}W^{-}\rightarrow 2l4j{\displaystyle{\not }}E_{T}$\protect & $24$\\\hline
\textbf{Backgrounds}\protect & \\
\hline
$t\bar{t}W\rightarrow WWW2j\rightarrow 2l4j{\displaystyle{\not }}E_{T}$\protect & $\sim 6.5\times 10^{3}$\\\hline
$HH\rightarrow WWWW\rightarrow 2l4j{\displaystyle{\not }}E_{T}$& $\sim 1.6\times 10^{3}$\\\hline
$WWW2j\rightarrow 2l4j{\displaystyle{\not }}E_{T}$\protect & $\sim 253$\\\hline
$HW2j\rightarrow WWW2j\rightarrow 2l4j{\displaystyle{\not }}E_{T}$\protect & $\sim 100$\\\hline
$HWZ\rightarrow WWW2j\rightarrow 2l4j{\displaystyle{\not }}E_{T}$\protect & $\sim 1.5$\\\hline
$HWW\rightarrow WWW2j\rightarrow 2l4j{\displaystyle{\not }}E_{T}$\protect & $\sim 6$\\\hline
$WWWW\rightarrow 2l4j{\displaystyle{\not }}E_{T}$& $\sim 3$\\\hline
\end{tabular}\caption{Number of same-sign di-lepton events and estimation of the corresponding backgrounds at the LHC for $\sqrt{S}=14$ TeV and $\int\mathcal{L}dt=100$ fb$^{-1}$. The signal corresponds to the case $M_{V}=500$ GeV, $M_{h}=125$ GeV, $m_{t}=171.3$ GeV, $a=1/4$, $d=1$ and $G_{V}=\sqrt{5}v/4$.}\label{bk2levents}
\end{table}%\newline
% \newline
\begin{table}[htb!]
\centering
\small\begin{tabular}[c]{|c|c|}
\hline
\textbf{Signal}\protect & \textbf{Number of Events}\\
\hline
$pp\rightarrow V^{0}V^{0}\rightarrow W^{+}W^{-}W^{+}W^{-}\rightarrow 3l2j{\displaystyle{\not }}E_{T}$\protect & $9$\\\hline
\textbf{Backgrounds}\protect & \\
\hline
$t\bar{t}W\rightarrow WWW2j\rightarrow 3l2j{\displaystyle{\not }}E_{T}$\protect & $\sim 2\times 10^{3}$\\\hline
$HH\rightarrow WWWW\rightarrow 3l2j{\displaystyle{\not }}E_{T}$& $\sim 10^{3}$\\\hline
$WZZ\rightarrow 3l2j{\displaystyle{\not }}E_{T}$& $\sim 10^{2}$\\\hline
$WWW2j\rightarrow 3l2j{\displaystyle{\not }}E_{T}$& $\sim 80$\\\hline
$HW2j\rightarrow WWW2j\rightarrow 3l2j{\displaystyle{\not }}E_{T}$\protect & $\sim 30$\\\hline
$HWW\rightarrow WWW2j\rightarrow 2l4j{\displaystyle{\not }}E_{T}$\protect & $\sim 4$\\\hline
$WWZZ\rightarrow 3l2j{\displaystyle{\not }}E_{T}$& $\sim 2$\\\hline
$WWWW\rightarrow 3l2j{\displaystyle{\not }}E_{T}$& $\sim 2$\\\hline
$HWZ\rightarrow WWW2j\rightarrow 3l2j{\displaystyle{\not }}E_{T}$\protect & $\sim 0.5$\\\hline
\end{tabular}\caption{Number of tri-lepton events and estimation of the corresponding backgrounds at the LHC for $\sqrt{S}=14$ TeV and $\int\mathcal{L}dt=100$ fb$^{-1}$. The signal corresponds to the case $M_{V}=500$ GeV, $M_{h}=125$ GeV, $m_{t}=171.3$ GeV, $a=1/4$, $d=1$ and $G_{V}=\sqrt{5}v/4$.}\label{bk3levents}
\end{table}%\newline
These numbers of multilepton events are comparable to those obtained from the decay of composite vector pairs produced via Drell-Yan anihilation. The numbers of multilepton events from the decay of composite vector pairs produced via Drell-Yan anihilation, which are independent on the $G_V$ coupling, are given in \cite{Barbieri:2010}. Since the vector pair production cross sections via the gluon fusion mechanism have a quadratic dependence on $d^2$, deviations of the parameter $d$ from $d=1$ will lead to a significant increase on the numbers of multilepton events.\newline
Tables \ref{bk2levents} and \ref{bk3levents} show the numbers of same-sign di-lepton and tri-lepton events and the estimation of the corresponding backgrounds. These backgrounds were computed using ALPGEN. The signals corresponding to same-sign di-lepton and tri-lepton events are hidden from the large backgrounds $t\bar{t}W$ and $HH$, whose corresponding number of events are two orders of magnitude larger than those corresponding to the signals. Therefore, the signals are very difficult to detect at colliders, since they are very suppressed by a factor $10^{-2}$ with respect to the $t\bar{t}W$ and $HH$ backgrounds. This explains why a heavy composite vector pair has not been seen at the LHC.\newline
Kinematical cuts on final-state leptons and jets have to be imposed to reduce the backgrounds. This will require interfacing ALPGEN to HERWIG which will provide an analysis of the full final state including a high cut on the scalar sum, $H_t$, of all the transverse momenta and of the missing energy in each event. It would also be interesting to extend the model by including a fourth quark generation and vector-like quarks and study their effects on the vector pair production at the LHC. This is beyond the scope of this work and is left for future studies. 
%To reduce the backgrounds, one has to impose kinematical cuts on the final state particles of the processes corresponding to the backgrounds. 
\section{Summary and conclusions}\label{sec7}
%The seach of heavy vector pair via the gluon fusion channel production might be feasible
In the framework of strongly interacting dynamics for EWSB, composite light scalar singlet and triplet heavy vector resonances may exist, and the interactions among themselves and with the Standard Model fermions and gauge bosons can be described by a $SU(2)_L\times SU(2)_R/SU(2)_{L+R}$ effective chiral Lagrangian. In this framework, the squared gluon fusion vector pair production amplitudes summed over the polarization and color states have been computed by considering that their only relevant contributions arise from the top quark in the triangular loop followed by a scalar propagator. The asymptotic behavior of the gluon fusion vector pair production amplitudes goes as $\frac{s}{M^2_V}$ at high energies and will have to be improved by the inclusion of a scalar-vector mixing term, with appropiate coupling. The gluon fusion vector pair production amplitudes depend on the couplings $c$, $d$, $g_V$ and on the masses $M_h$, $M_V$ and $m_t$. The unitarity constraints in the elastic channel of longitudinal gauge boson scattering and in the inelastic scattering of two longitudinal SM gauge bosons into SM fermions pairs, determine the relevant parameter space. A discussion about the phenomenology of the composite vector pair production via the gluon fusion mechanism at the LHC, has been presented. For a vector mass between $500$ GeV and $1$ TeV, for $M_h=125$ GeV and $m_{t}=171.3$ GeV, the total cross sections for the production of the $V^{+}V^{-}$ and $V^{0}V^{0}$ final states at the LHC by the gluon fusion mechanism have been computed. These total cross sections are of the order of few $fb$. It is of remarkable relevance that the only process which produces a $V^{0}V^{0}$ final state is the gluon fusion mechanism, since the $V^{0}V^{0}$ final state is absent in the Drell-Yan process. The $V^{+}V^{-}$ final-state production cross section via the gluon fusion mechanism is comparable with the $V^{+}V^{-}$ Drell-Yan production cross section.
%The neutral composite vector pair production final state is the only
The $V^{0}V^{0}$ production total cross section via the gluon fusion mechanism is $4$ times larger than the $V^{+}V^{-}$ production cross section since the ratio between the symmetry factors for the $hV^{0}V^{0}$ and $hV^{+}V^{-}$ vertex is equal to 2. These total cross sections can be strongly increased since they depend quadratically on the scalar-vector coupling $d$. The expected same-sign di-lepton and tri-lepton events are of order of $10$ for an integrated luminosity of $100$ fb$^{-1}$ and are two orders of magnitude lower than the large backgrounds $t\bar{t}W$ and $HH$. The signals corresponding to the same-sign di-lepton and tri-lepton events are suppressed by two orders of magnitude with respect to the $t\bar{t}W$ and $HH$ backgrounds and are therefore difficult to detect at the LHC. Kinematical cuts on final-state leptons and jets have to be imposed to reduce the backgrounds. A detailed investigation of the SM backgrounds, wherein acceptance cuts on final-state leptons and jets, as well as detector effects, are expected to play a role, is deferred to future work. Other possible direction for future work along these lines would be to study the composite vectors effects in the $h\to\gamma\gamma$ decay to determine the restrictions that the aforementioned $SU(2)_L\times SU(2)_R/SU(2)_{L+R}$ Effective Lagrangian should have, in order to explain the excess of events in this decay, recently observed at the LHC. An extension of the model would include a fourth quark generation and/or vector-like quarks as well as direct couplings between heavy quarks and composite vectors. Their effects on the vector pair production and on the $h\to\gamma\gamma$ decay at the LHC may be useful to study. To address all these issues requires careful investigations that are beyond the scope of this work. They are left for future studies.

\subsection*{Acknowledgements}
The author is greatly indebted to Professors Riccardo Barbieri, Gino Isidori, Alfonso Zerwekh and Claudio Dib for many useful suggestions and for careful reading of the manuscript. The author also thanks his wife for drawing the Feynman diagrams.

%\newpage

\end{document}